\documentclass[fleqn,10pt]{wlscirep}
\usepackage[utf8]{inputenc}
\usepackage[T1]{fontenc}

\usepackage{multirow}
\usepackage[normalem]{ulem}
\useunder{\uline}{\ul}{}
\usepackage{graphicx}
\usepackage{url}
\usepackage{hyperref}
\title{EXACT: A collaboration toolset for algorithm-aided annotation of images with annotation version control}
\author[1,2,*]{Christian Marzahl}
\author[1,4]{Marc Aubreville}
\author[3]{Christof A. Bertram}
\author[1]{Jennifer Maier}
\author[1]{Christian Bergler}
\author[2]{Christine Kr{\"o}ger}
\author[2]{J{\"o}rn Voigt}
\author[1]{Katharina Breininger}
\author[3]{Robert Klopfleisch}
\author[1]{Andreas Maier}

\affil[1]{Pattern Recognition Lab, Friedrich-Alexander-Universität Erlangen-Nürnberg, Erlangen, Germany}
\affil[2]{Research and Development, EUROIMMUN Medizinische Labordiagnostika AG, Lübeck, Germany}
\affil[3]{Institute of Veterinary Pathology, Freie Universität Berlin, Germany}
\affil[4]{Faculty of Computer Science, Technische Hochschule Ingolstadt, Ingolstadt, Germany}

\affil[*]{c.marzahl@euroimmun.de}
\begin{abstract}

In many research areas, scientific progress is accelerated by multidisciplinary access to image data and their interdisciplinary annotation. However, keeping track of these annotations to ensure a high-quality multi-purpose data set is a challenging and labour intensive task. 
We developed the open-source online platform EXACT (EXpert Algorithm Collaboration Tool) that enables the collaborative interdisciplinary analysis of images from different domains online and offline. EXACT supports multi-gigapixel medical whole slide images as well as image series with thousands of images. The software utilises a flexible plugin system that can be adapted to diverse applications such as counting mitotic figures with a screening mode, finding false annotations on a novel validation view, or using the latest deep learning image analysis technologies. This is combined with a version control system which makes it possible to keep track of changes in the data sets and, for example, to link the results of deep learning experiments to specific data set versions.
EXACT is freely available and has already been successfully applied to a broad range of annotation tasks, including highly diverse applications like deep learning supported cytology scoring, interdisciplinary multi-centre whole slide image tumour annotation, and highly specialised whale sound spectroscopy clustering.
\end{abstract}
\begin{document}

\flushbottom
\maketitle
\thispagestyle{empty}
\section*{Introduction}

The joint interdisciplinary evaluation of images is critical to scientific progress in many research areas. 
Specialised interpretation of images strongly benefits from cross-discipline cooperation among experts from different disciplines such as the annotation of pathology microscopy slides with the aim of facilitating routine pathology tasks. The strenuous annotation work can be greatly simplified by customised algorithmic support for medical experts provided by engineers and computer scientists. However, this interdisciplinary cooperation has specific demands on all parties involved. One important aspect to be observed is data privacy and protection. Regulations must be put in place to control who is allowed to access which image set and which data are shared. Furthermore, the tools for viewing and annotating images must be efficient and user-friendly in order to achieve a high level of acceptance among medical professionals. Computer-scientists, however, require traceable high-quality and high-quantity data sets which are essential for reproducibility when creating accurate machine learning algorithms.   
In order to meet these diverse requirements for annotating image data, a wide variety of open-source software solutions have been designed and published in recent years. These software solutions can be divided into three groups: firstly, offline annotation tools like SlideRunner \cite{aubreville2018sliderunner}, AnnotatorJ \cite{hollandi2020annotatorj}, Icy \cite{de2012icy}, or QuPath \cite{bankhead2017qupath}. Secondly, web-based solutions focusing on cooperation like Cytomine \cite{maree2016collaborative} or OpenHI \cite{puttapirat2018openhi}. And finally, platforms that combine established solutions like Icytomine \cite{obando2019icytomine} which combines both Icy and Cytomine. All these solutions support whole slide images (WSIs) and provide open-source access for scientific research purposes.

\begin{table}[]
    \centering
    \caption{Features of EXACT regarding applications, data set annotation and machine learning.}
    \label{tabEXACTKeyFeatures}
    %\resizebox{\textwidth}{!}
    {%
        \begin{tabular}{cc|l}
        \multicolumn{2}{c|}{Features}                          & Description \\ \hline
        \multirow{14}{*}{\rotatebox[origin=c]{90}{Application}} 
                             & Online                  &   EXACT is a Django-based server application with a browser client    \\
                             & Cross-platform          &   The server can be installed on Windows, Linux, Mac and all other systems      \\
                             &                         &   with Docker support      \\
                             & Multi-Center            &   The multi-center support allows sharing data across multiple institutes with appropriate       \\
                             &                         &               data privacy management       \\
                             & REST-API                &   The REST-API supports language independent create, read, update and delete (CRUD) \\
                             &                         &   operations on all database fields including image upload and download\\
                             & Language                &   The web-server is written in Python while the web-client is HTML and JavaScript based\\
                             & Plugins                 &   Allow the frontend and backend integration of domain-specific features and analyses \\
                             & Image-set administration&   Combining images to a folder like structure with team access rights and     \\
                             &                         &   shared annotations scheme     \\
                             & File Formats            &   Images: .tif, .png, bmp, .jpeg, .dcm (partially), .webp        \\
                             &                         &   WSI scanner: .svs, .vms, .vmu, .ndpi, .scn, .mrxs, .iSyntax, .svslide,  .bif, .czi, .tif      \\
                             &                         &   Time-Series: .avi, .mkt, .tif        \\
                             & User management         &   Individual user or group rights for CRUD operations on the database \\ \cline{2-3}
        \multirow{4}{*}{\rotatebox[origin=c]{90}{Annotation}}  
                             & Types                   &   Box, polygon, line, circle and per image (classification) annotations\\
                             & Templates               &   Define a framework for general annotation properties like type (box, polygon etc.), \\
                             &                         &   colour and default size\\
                             & Single click            &   Single click annotations with background knowledge provied by the templates\\
                             & Guided screening        &   A persistent screening mode in a user defined resolution which saves the progress\\ \cline{2-3}
        \multirow{3}{*}{\rotatebox[origin=c]{90}{ML}}             
                             & Version control         &   Enables the versioning of image sets with the corresponding image list and annotations  \\  
                             & Inference               &   Performing inference on client side via browser, on server side via Python or over the \\ 
                             &                         &   REST-API  \\
    \end{tabular}%
    }
\end{table}

In the following, we define a set of specific requirements for collaborative annotation software that are - in this combination and at the time of this publication - not satisfied in open-source solutions. Furthermore, we introduce annotation templates and annotation versioning as new requirements.

The software should be usable online and offline, while providing multi-centre support for interdisciplinary cooperation and an easy-to-use API to facilitate integration with existing software. Furthermore, an extensible plugin system for easy adaptation to specific use cases should be included and an image-set administration aspect to manage and group images with restricted access through a user management system. Bounding boxes and polygon annotations as well as single click support are critical features for an efficient and flexible annotation workflow. Annotation templates enforce a unique naming scheme essential for standardisation and allow the incorporation of background knowledge. Additionally, guided screening to annotate WSIs systematically should be supported. Finally, to achieve reproducible results in the machine learning algorithm development process, a version control system for annotations and the possibility to perform inference of deep learning models is advantageous. %These specific requirements for collaborative annotations have not been incorporated into a open-source solution. 
Based on these requirements, we introduce EXACT, a novel online open-source software solution for massive collaboration in the age of deep learning and big data. EXACT was developed with seamless interaction to offline clients in mind, and interoperates with the established SlideRunner software \cite{aubreville2018sliderunner}.   

% TODO: Spelling check
In the following section, we describe the architecture of EXACT with its key features (see Table~\ref{tabEXACTKeyFeatures}) and the design principles behind them. In the chapter "EXACT's applications", we showcase four very different projects where EXACT was applied to create high-quantity and high-quality data sets. Finally, we present a discussion and outlook.

\begin{figure}[hbt!]
\includegraphics[width=1\textwidth]{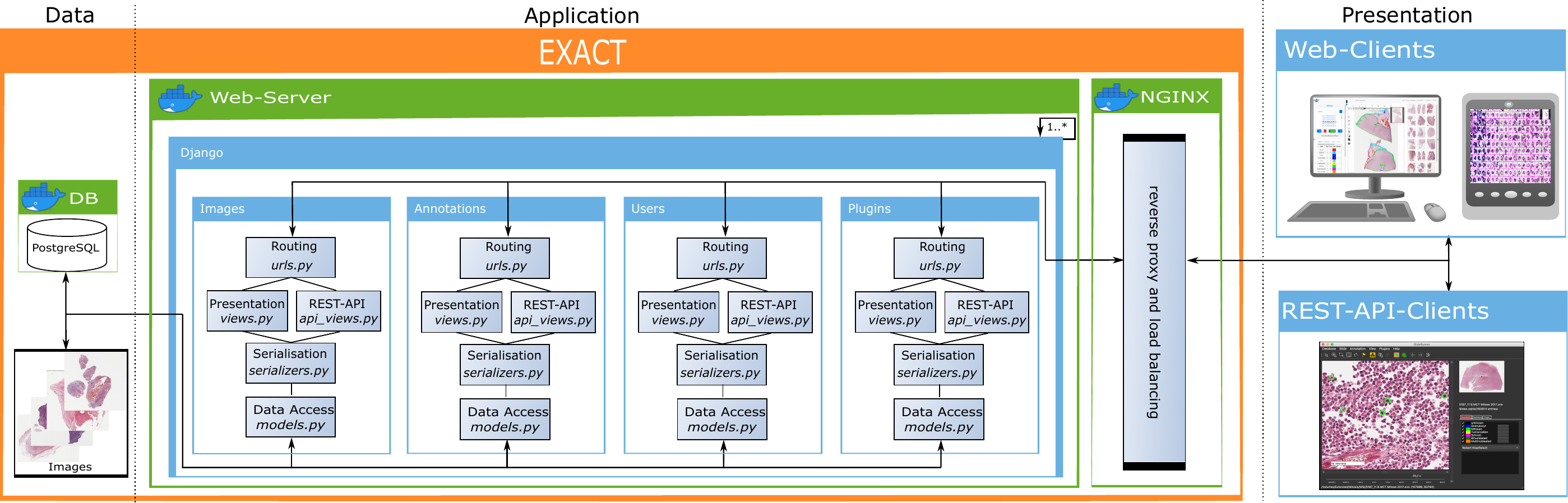}
\caption{EXACT's three-tier architecture. Left: The data tier contains the PostgresSQL Docker container and the images, which can be saved within the docker container or on the file system. Center: The application tier with the web-server Docker container instantiating Django instances with the corresponding modules. These modules handle images, annotations, users and plugin requests from the presentation tier and access the data tier to retrieve the stored information. NGINX works as a reverse proxy and handles EXACT's load balancing. Right: The presentation tier contains the EXACT web client or third-party applications like SlideRunner, which send requests via the provided REST-API.}
\label{figArchitektur}
\end{figure}

% TODO: Spelling check
\section*{EXACT's architectural design and features}

%EXACT substantially extends the established online open-source software ImageTagger \cite{fiedler2018imagetagger}, which was developed for the RoboCup competition to create training data for machine learning projects. We chose ImageTagger because it already fulfilled many of our basic requirements and %, in contrast to the much more complex Cytomine software, 
%allowed due to its reduced complexity, fundamental changes to the software design which were necessary to integrate functions like image set versioning.

The development of EXACT was based on the established online open-source software ImageTagger \cite{fiedler2018imagetagger}, which was developed for the RoboCup competition to create training data for machine learning projects. It already fulfils many of our basic requirements. Due to its low complexity, it allows for fundamental changes to the software design which are necessary to integrate functions like image set versioning.
ImageTagger uses Django as its web framework, a Postgres database system and hypertext markup language (HTML) with JavaScript as frontend user interface. The following basic features and modules are substantially extended from or added to ImageTagger: We have added the docker encapsulation, implemented the complete REST-API and have changed the image viewer to support the open-source software OpenSeadragon, which provides functionality to view WSIs in the browser. In this context, we have extended the images module to handle WSIs and provide functions to convert images into compatible WSI formats. Furthermore, we have made many performance adjustments to transfer annotations in parallel, display multiple annotation types simultaneously and synchronise annotations of other users. Also, we have completely redesigned the image viewer to display thumbnails of the image set and have created the possibility to include plugins. 
In the following subsections, we will first describe the architecture including the application and presentation tier. This is continued by introducing additional aspects of this software and their specialised extensions, like inference, data privacy, annotation maps, image set versioning, crowd-sourcing and annotation templates. Further implementation details are provided via videos, jupyter notebooks or setup and code files in the supplementary information section (Tab. \ref{tabDocumentation}). 

\subsection*{Architecture}

%In this section, we first describe EXACT's architecture followed by novel features which are not only tailored for pathologists and researchers working with WSIs but can also be applied to 2D image data in general. 

EXACT supports Docker to facilitate deployment and to enable a wide range of installation scenarios ranging from single-user, single-computer setups to massive cloud deployment with modern load balancing mechanisms (see Supplementary Video S10).  
% Plugins % REST-API % Diagramm  Installation.mp4
EXACT is designed as a three-tier architecture containing the data, application, and presentation tier (Fig. \ref{figArchitektur}). While the data and application tier are capsuled within Docker containers, the presentation tier is executed at the client side in HTML and JavaScript. This tier-based approach supports the development of secure applications by enforcing clearly defined interfaces between tiers and ensures that data access pipelines can not bypass tiers. The data tier includes a Postgres database system and the uploaded images and provides its content exclusively to the application tier. 

\subsubsection*{Application tier}
  
The application tier accesses the data tier to save information and to provide it to the presentation tier via a REST-API or as rendered HTML pages. EXACT uses Django as its web framework with four main modules (see Fig. \ref{figArchitektur}, namely the \href{https://github.com/ChristianMarzahl/Exact/tree/master/exact/exact/images}{images}, \href{https://github.com/ChristianMarzahl/Exact/tree/master/exact/exact/users}{users}, \href{https://github.com/ChristianMarzahl/Exact/tree/master/exact/exact/annotations}{annotations}, and \href{https://github.com/ChristianMarzahl/Exact/tree/master/exact/plugins}{plugins} modules). Each module is responsible for one group of tasks and is as independent as possible from the other modules. All modules implement functions for saving information to the database or file system and for creating HTML views. Furthermore, the modules define how to serialise data and provide a REST-API and a route request. 

\begin{figure}[hbt!]
\includegraphics[width=1\textwidth]{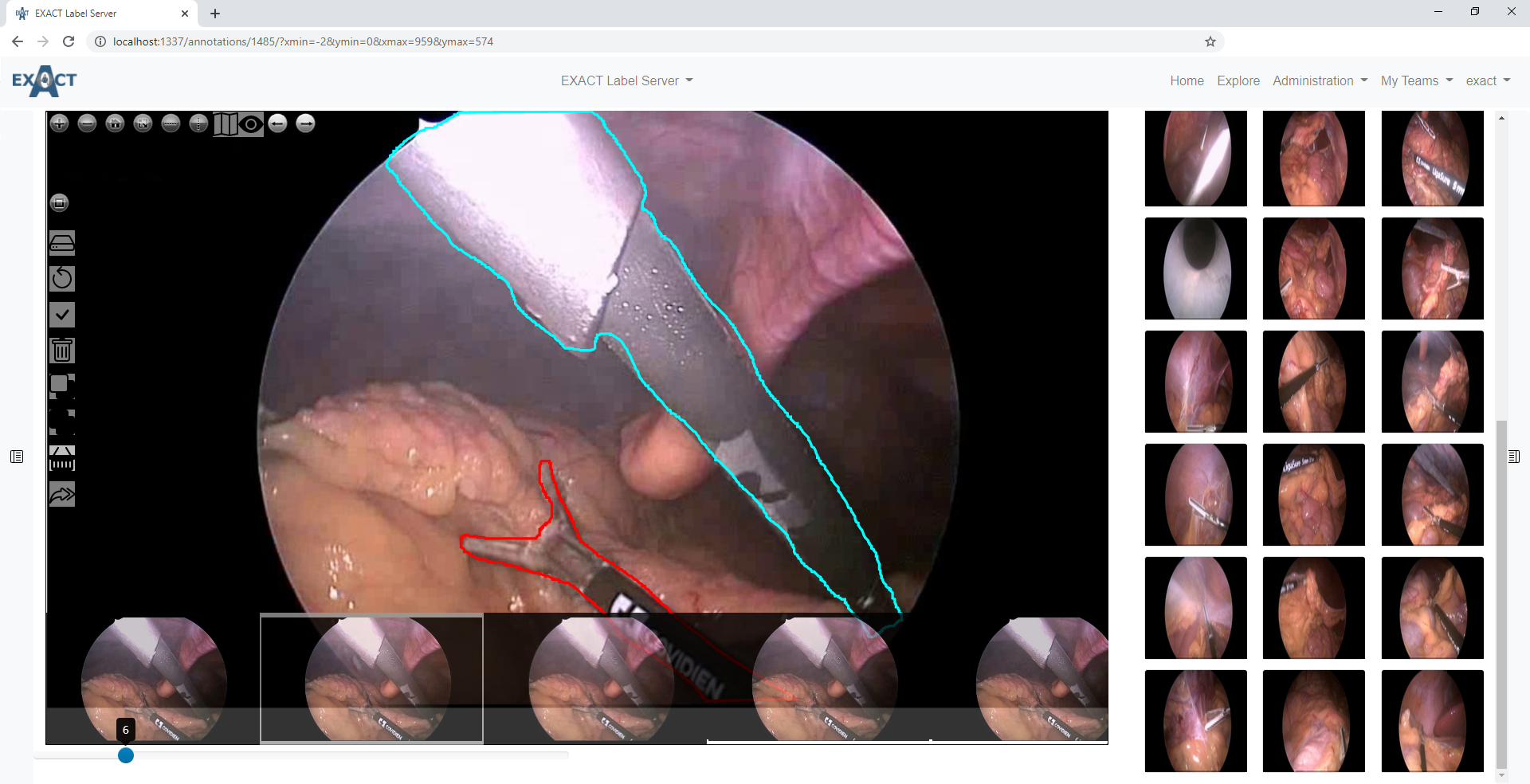}
\caption{Left: An example frame from a laparoscopic colorectal video with annotated surgical instruments. Below the current frame (6), small previews of the previous and the next frames are displayed. Right: A browsing view to provide an overview of different image sets from the Robust Endoscopic Vision Challenge 2019~\cite{maier2021heidelberg,ross2021comparative}.}
\label{figFrameExampleView}
\end{figure}

The \href{https://github.com/ChristianMarzahl/Exact/tree/master/exact/exact/images}{images} module is responsible for all image-based create, read, update and delete (CRUD) operations, and provides the logic to save all supported image formats and to provide them as a complete image or in a tile-based manner for WSIs. This multi type image support is implemented by converting all uploaded images that are not compatible with OpenSlide~\cite{goode2013openslide} into an OpenSlide compatible format, if supported, and saving them as an image pyramid. The formats and scanners that are supported by OpenSlide or our converter pipeline is listed in table \ref{tabEXACTKeyFeatures}. %: Aperio (.svs, .tif), Hamamatsu (.vms, .vmu, .ndpi), Leica (.scn), MIRAX (.mrxs), Philips (.tiff, .iSyntax), Sakura (.svslide), Trestle (.tif), Ventana (.bif, .tif), Generic tiled TIFF (.tif), Zeiss (.czi), JPEG, PNG and WebP.
EXACT's open-source codebase allows developers to extend the list of supported image formats to their requirements and image dimensions. An example for multi-dimensional image data support is the audio video interleave (.avi) format. To support videos, EXACT converts each frame and handles the set of images as a individual WSIs with OpenSlide (Fig. \ref{figFrameExampleView}). Additionally, the \href{https://github.com/ChristianMarzahl/Exact/tree/master/exact/exact/images}{images} module contains the image sets functionality which basically act as folders for the images and are assigned to teams to monitor user access rights. The Supplementary Videos (S7, S8) describe the creation of image sets and the upload of images.%\url{https://youtu.be/F3lV-IvT1M4}.

The \href{https://github.com/ChristianMarzahl/Exact/tree/master/exact/exact/annotations}{annotation} module is responsible for all CRUD operations regarding annotations, verification, media files and the annotation versioning system. The annotation model saves annotation information about the annotation type, image, the creator and last editor with time stamps, JSON based meta data and the vector of coordinates to the database. The vector information is saved as JSON and contains the image coordinates of the annotation. The advantage of using JSON to store coordinates is the ability to search for annotations using vector coordinates in SQL. %(\textit{SELECT * FROM annotation where  CAST(vector ->> 'x1'  AS INTEGER) > 5 and CAST(vector ->> 'x1' AS INTEGER) < 100;}) 
Furthermore, JSON provides the flexibility to adapt the representation of the vector to the target image format and dimensions.  

The \href{https://github.com/ChristianMarzahl/Exact/tree/master/exact/plugins}{plugin} module handles analysis or visualisation plugins which are specialised for specific research questions or data sets. One of these plugins is a persistent user-based screening mode which enables the user to systematically screen a WSI or parts of it on a self-defined zoom level (Fig. \ref{figAnnotationView}). This plugin, which is crucial to create high quality data sets, is implemented and used in the following manner: The user defines a zoom level and the algorithm divides the WSI in equal-sized patches with an overlap of 15\% and saves the calculated screening map to the database. While the user is screening the WSI, the progress is constantly visualised at a thumbnail view of the WSI and the user's position on the WSI is saved to the database to recover the position if the screening has to be continued later. 

The \href{https://github.com/ChristianMarzahl/Exact/tree/master/exact/exact/users}{users} module handles the CRUD operations for users and teams, and it further manages the user access rights. It is therefore involved in every server request to check if the request has the necessary CRUD rights (see Supplementary Video S14). To keep the annotations consistent, deleted users are anonymised and deactivated while their annotations are left unchanged. 

An additional module is the \href{https://github.com/ChristianMarzahl/Exact/tree/master/exact/exact/datasets}{data sets} module. It provides features to automatically download and setup predefined data sets with their annotations from the EXACT user interface. The list of available data sets can be extended by adding an HTML template, which provides background information like the number of images or the data set source, and by implementing a download and setup function (see Supplementary Video S5). %A demonstration video can be viewed at \textit{Demo Dataset.mp4}. %\url{https://youtu.be/hi23nhz0rWQ}.

\subsubsection*{Presentation tier}

\begin{figure}[hbt!]
\includegraphics[width=1\textwidth]{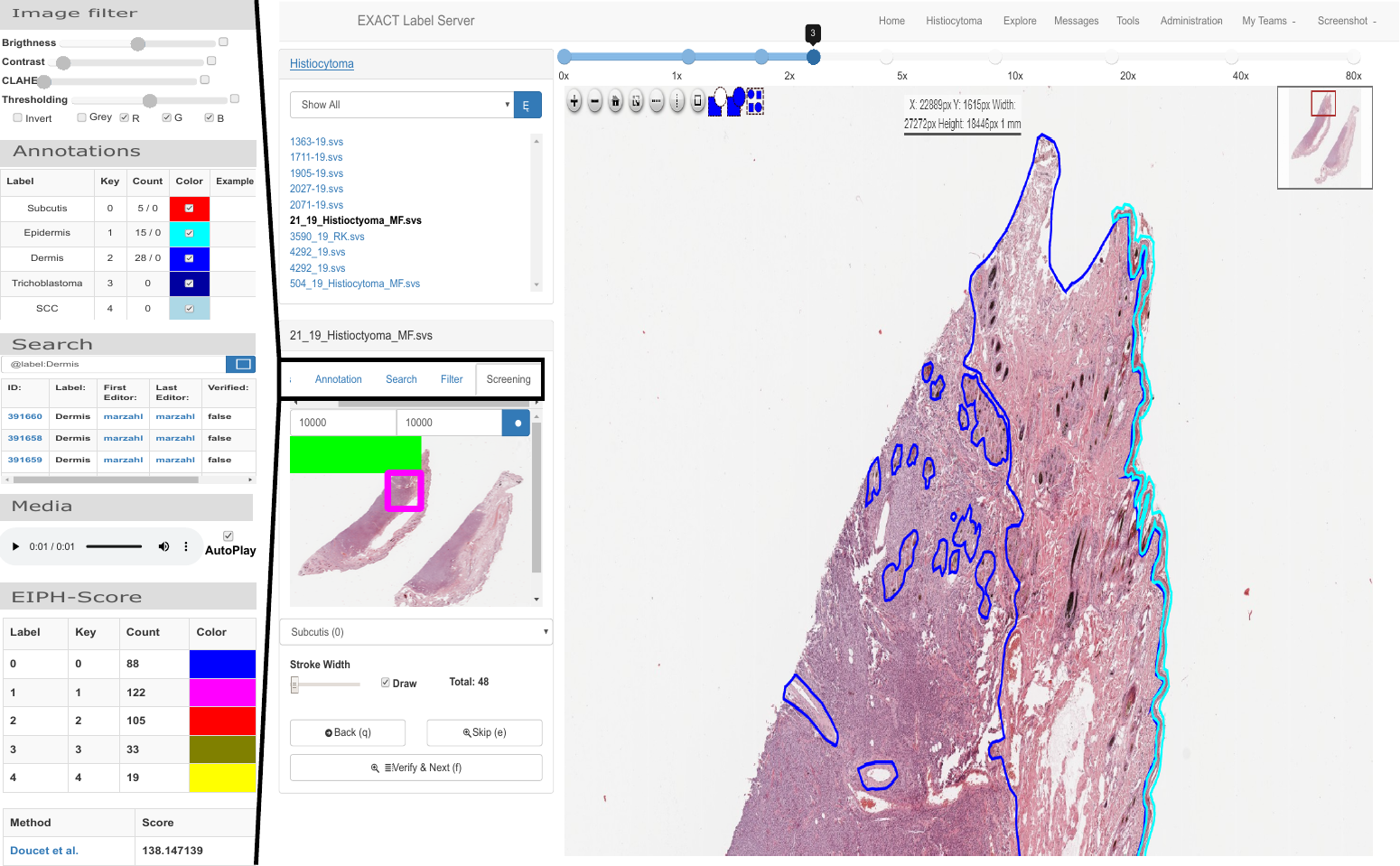}
\caption{Left: Five examples of plugins (from top to bottom): The image filter plugin allows to make common intensity adjustments to the image, the annotation plugin shows the available annotations and their frequency of use. The search field allows to query the database for arbitrary annotation properties. The media plugin can be used to play media files attached to an annotation. The EIPH-Score plugin is an example of a domain-specific plugin, allowing to calculate the Doucet score\cite{doucet2002alveolar}. Right: A screenshot of the annotation view depicting a WSI with polygon annotations, the list of images in the image set and the screening mode plugin, which enables the user to screen the image persistently. The screening plugin visualise the screened area in green and a purple rectangle for the current field of view. }
\label{figAnnotationView}
\end{figure}

The presentation tier is programmed in HTML and JavaScript. All dynamic web-page contents like annotations, images or sub-images (tiles) for WSIs are loaded via JavaScript over the REST-API. The pagination-based REST-API implementation allow to load information chunk-wise from the server and therefore enable the transfer of huge quantities of data (e.g., hundreds of thousands of annotations per WSI) in parallel. We incorporated the open-source software OpenSeadragon as JavaScript-based image viewer with WSI support. A visualisation of the presentation tier is shown in Fig. \ref{figAnnotationView}.

\subsubsection*{Inference}

Different modes for inference of deep learning models are supported to match the requirements across different use cases. In general, the inference can be performed directly on the server. For applications that require fast response times, the execution of JavaScript-based TensorFlow models is implemented by initially transferring the deep-learning model for the corresponding modality from the server via the REST-API to the JavaScript client. Afterwards, the model is executed on the current field of view of the image. The resulting annotations can then be rejected or confirmed and transferred to the server. 
For high-throughput applications, the inference load can be distributed on multiple machines by downloading the model and the WSIs via the REST-API and synchronising the results after performing inference. An inference example for equine asthma cytology images can be accessed at \href{https://nbviewer.jupyter.org/github/ChristianMarzahl/Exact/blob/master/doc/Inference Asthma.ipynb}{doc/Inference Asthma.ipynb} or as Supplementary Video S9. %\url{https://youtu.be/xP4YAp678EM}.

%\begin{figure}[hbt!]
%\includegraphics[width=1\textwidth]{FrameExample.png}
%\caption{Left: An example frame from a contrast angiography of %right forearm with the radius (left bone) and the ulna (right bone) %in the lower two thirds of the picture and part of the humerus on %the top part op the picture. At the bottom of the frame the %intra-vascular catheter is clearly visible and annotated with a %bounding box. Below the current frame (27) are next and last frames %visible as selectable previews. Right: A preview of the other %imageset videos.}
%https://exact.cs.fau.de/annotations/5017
%\label{figFrameExampleView}
%\end{figure}

\subsection*{Data privacy and multi-centre support}

Medical data should naturally be subject to the highest safety standards possible. Despite that, in order to enable interdisciplinary medical research and cooperation between different groups and locations, it can be necessary to share medical image data anonymously and in strict consideration of data privacy.
Therefore, EXACT ensures the original image data, which may contain patient information (file name, metadata) to remain within the original institution while the actual data exchange between experts and institutes is executed on small sub-images via decentralised image storage. 
%while the cooperation between experts and institutes is facilitated by the use of decentralised image storage.
Technically this was implemented in several steps. Firstly, all server communication is protected with Hypertext Transfer Protocol Secure (HTTPS) and access is restricted via a user authentication system. Secondly, when transferring the images to an EXACT server instance, a new private name derived from the file name and a pseudonymised public name is generated. The pseudonymised public name is generated by the current date-time followed by a four-digit hash function of the original name (yymmdd-hhmm-****). Thirdly, for cooperation between different institutes, virtual image sets are supported. Here the information (for example annotations) is imported from several EXACT instances to a central server. However, access to the images themselves is always provided by the institute owning the data in compliance with their respective data privacy policy for images. This means that only the requested raw pixel data for the field of view is transferred to the collaborator, but not the image container or any metadata.

\subsection*{Annotation map screening mode}

\begin{figure}[hbt!]
\includegraphics[width=1\textwidth]{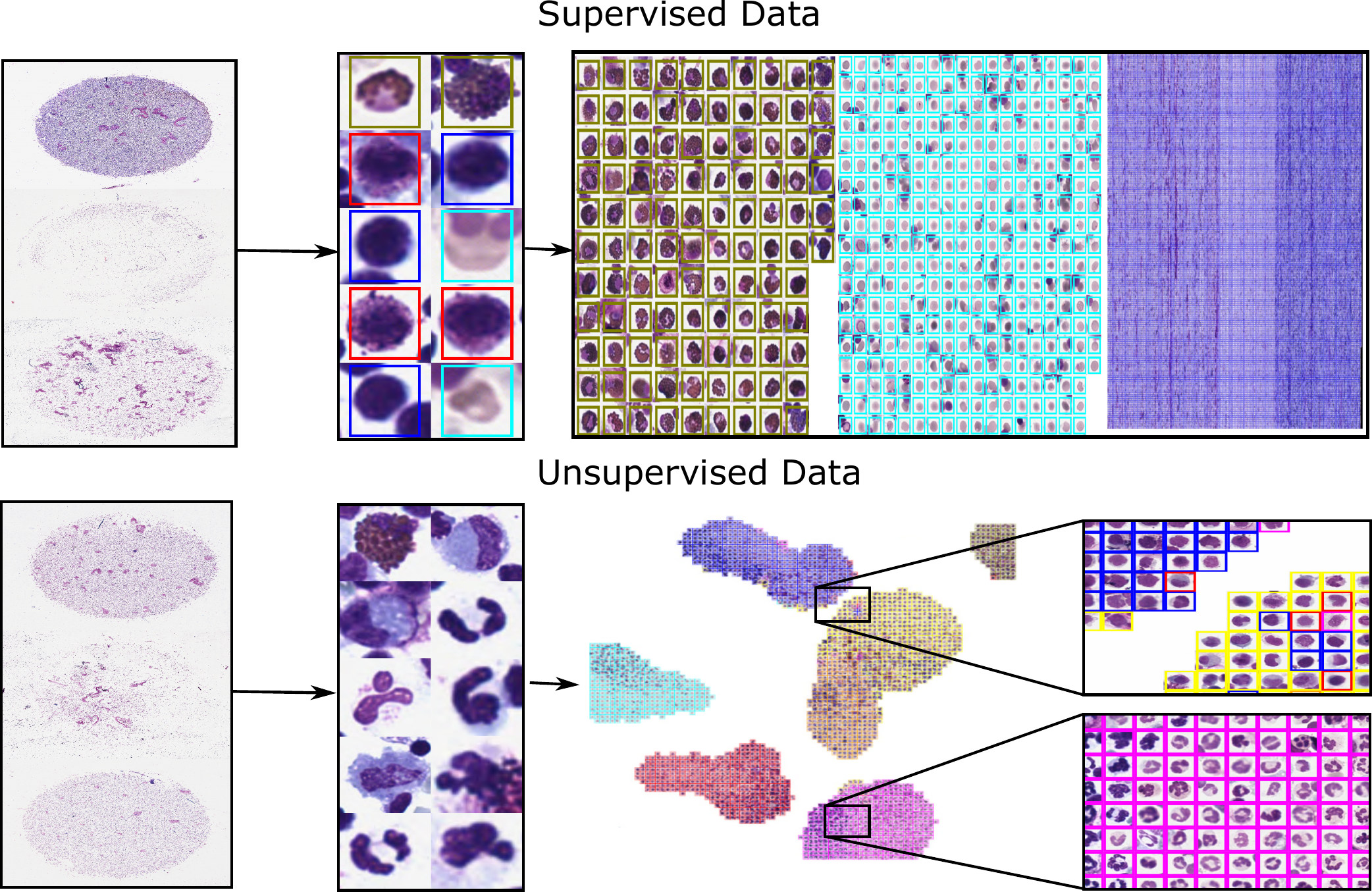}
\caption{Top row: Supervised single-cell validation, with annotation maps generated from three labelled equine asthma WSIs where each colour represents one class of cells. Bottom row:  UMAP \cite{2018arXivUMAP} dimensionality reduction approach in an unsupervised setting. The segmented equine asthma cells are first classified. Then, features for each cell are extracted. Afterwards, the high-dimensional features are transformed into a two-dimensional representation and visualised in a new image. Both approaches allow the user to verify and enhance the automatic classification results.  }
\label{figAnnotationMap}
\end{figure}

For applications that focus on annotation quality \cite{aubreville2018sliderunner,obando2019icytomine,fiedler2018imagetagger}, a specialised validation mode is implemented that allows for a verification of each individual annotation.
%A specialised validation mode where each annotation can be verified individually is standard for this type of application \cite{aubreville2018sliderunner,obando2019icytomine,fiedler2018imagetagger}. 
For data sets with hundreds or thousands of annotations, this is an important but error-prone, labour-intensive and time-consuming task. This becomes even more complicated for usage scenarios where each cell can receive multiple labels by one or multiple users. To make this validation process more convenient, we propose so-called annotation maps which can be efficiently processed using the screening mode. Annotation maps visualise all annotations belonging to one label in a matrix-like fashion which makes it easy to identify outliers. For efficient handling, a new image is created for each class which consists of all corresponding annotations which can then be viewed in the screening mode (Fig. \ref{figAnnotationMap} top and Supplementary Video S2). The annotation maps can be efficiently screened for errors, while the users can define how many annotations they want to see simultaneously. Corrections made on these screening images are synchronised with the original data. 
% Done: Annoation Map Video with creation

An advanced extension of this method is the clustering of labelled and unlabelled images or image patches. This manner of presentation allows the user to efficiently create initial labels or to quickly validate prior annotations, since similar images which are likely to have similar labels are displayed closely together. The clustering pipeline consists of three steps. Firstly, characteristic features are extracted from each image, for example, by deep learning or classic image processing. Secondly, the extracted high-dimensional features are transformed into two-dimensional features, for example, using t-SNE \cite{maaten2008visualizing}, PCA \cite{wold1987principal} or UMAP \cite{2018arXivUMAP}. Finally, the extracted image patches are drawn in a new image container according to their nearest two-dimensional feature representation, which does not overlay any other image patches. The resulting image is visualised for labelling or validation (Fig. \ref{figAnnotationMap} bottom) in EXACT. A detailed code example can be accessed at \href{https://nbviewer.jupyter.org/github/ChristianMarzahl/Exact/blob/master/doc/ClusterCells.ipynb}{doc/ClusterCells.ipynb} in combination with a Supplementary Video S4. %\url{https://youtu.be/Wvz-Nv4dNOE}. Asthma Clustering.mp4
% Done: Asthma Cluster Video with creation

\subsection*{Image set versioning and machine learning support}

In general, two main criteria in research and medical applications are reproducibility and traceability of results and experiments. Especially reproducibility is non-trivial in settings where researchers from different fields like medicine and computer science work together and make adjustments to data sets over time. In software development, it is an established process to use version control systems (such as git or subversion) for source code to coordinate the collaboration between software developers and keep code changes traceable. Remarkably, this process is to our knowledge not provided by any open source software for annotations on medical data sets. To implement this feature, we included a versioning system with functions that support traceability of annotations and attach experimental results to versions. If a version is added to a data set, the current annotation state, an optional description, and the current list of images in the data set is saved. If a user leaves a project, he or she is not deleted from EXACT but only deactivated and anonymised so that versioned annotations are not affected. In contrast, if an image is deleted from the image set, all annotations are lost due to the impracticability of versioning WSIs with multiple gigabytes of size. For example for training machine learning algorithms, the annotations can be filtered by versions and exported in user-defined text formats or per script using the provided REST-API. This supports the users to perform experiments on defined, reproducible data sets while providing the flexibility to export input data to a wide range of machine learning frameworks. 
Additionally, training artefacts like performance metrics, annotations, or generated models can be uploaded and attached to a version. In combination with the virtual image set function introduced previously in this article, it is possible to create virtual training, testing, and validation sets. This combination of versions and virtual image sets helps to keep track of different experiment versions and supports the comparability of results (see Supplementary Video S15). %\url{https://youtu.be/WeOWxXaYc0g}.
% Done: Video: Create imageset, change, track changes; upload model

\subsection*{Crowd-sourcing and study support}

One of the biggest challenges in developing, training, testing, and validating state-of-the-art machine learning algorithms is the availability of high-quality, high-quantity labelled image databases. Crowd-sourcing has numerous successful applications in the medical field \cite{orting2019survey} and crowd-algorithm collaboration has the potential to decrease the human effort \cite{marzahl2020crowd}. EXACT supports this development by providing multiple features for managing crowd-sourcing. Firstly, the user privilege system allows to set specific rights like annotation or validation to users or user groups. Secondly, the crowd- or expert-algorithm collaboration is assisted by importing pre-computed annotations or generating them on-premise with machine learning models. Finally, EXACT supports multiple annotation modes like:
\begin{enumerate}
    \item \textit{Cooperative}: One user can verify the image, and each user sees all other annotations.
    \item \textit{Competitive or Blind}: Every user must verify every image and can't see other users' annotations.
    \item \textit{Second opinion}: A predefined number of the users must verify every annotation. 
\end{enumerate}
% ToDo Optional: EIPH300 alle Anwender / Download annotations make simple statistics

\subsection*{Annotation templates}

Standardisation is critical to encourage cooperation, interoperability and efficiency. To support this, EXACT introduces annotation templates, which allow to define a set of properties of annotations associated with a defined label. Annotation templates contain general information about the target structure like a name, an example image, the sort order in which the annotation should be displayed on the user interface, display colour, keyboard shortcuts to efficiently assign the label to an annotation, and default size. Default sizes enable the user to introduce background knowledge into the annotation process; this allows for efficient single click annotations and reduces the need to further adjust annotations. One or more annotation templates are grouped to products with pieces of information like name or description and can be assigned to image sets. The products in turn can be assigned to multiple image sets and support the reproducibility of the annotation process by enforcing a standard naming and annotation schema (see Supplementary Video S3).% \url{https://youtu.be/4XdWLaqy9UA}.   Annotation Templates.mp4
% Done: Show annotation template 

\section*{EXACT's applications}

In the following sections, we present several previously published usage scenarios using EXACT and describe how they made use of EXACT's features to increase efficiency and annotation quality.
%The foundation for the use cases is that EXACT's source-code is freely available under open-source license at \url{https://github.com/ChristianMarzahl/Exact} together with a python REST-API implementation at \url{https://github.com/ChristianMarzahl/EXACT-Sync} for streamlined integration into existing projects. We additionally provide scripts to set up a local server with the virtualisation software Docker as well as example scripts for the most common EXACT use-cases like up- and downloading of data or creating cluster- or annotations maps at the \textit{doc/} folder of the repository.

\begin{figure}[hbt!]
\includegraphics[width=1\textwidth]{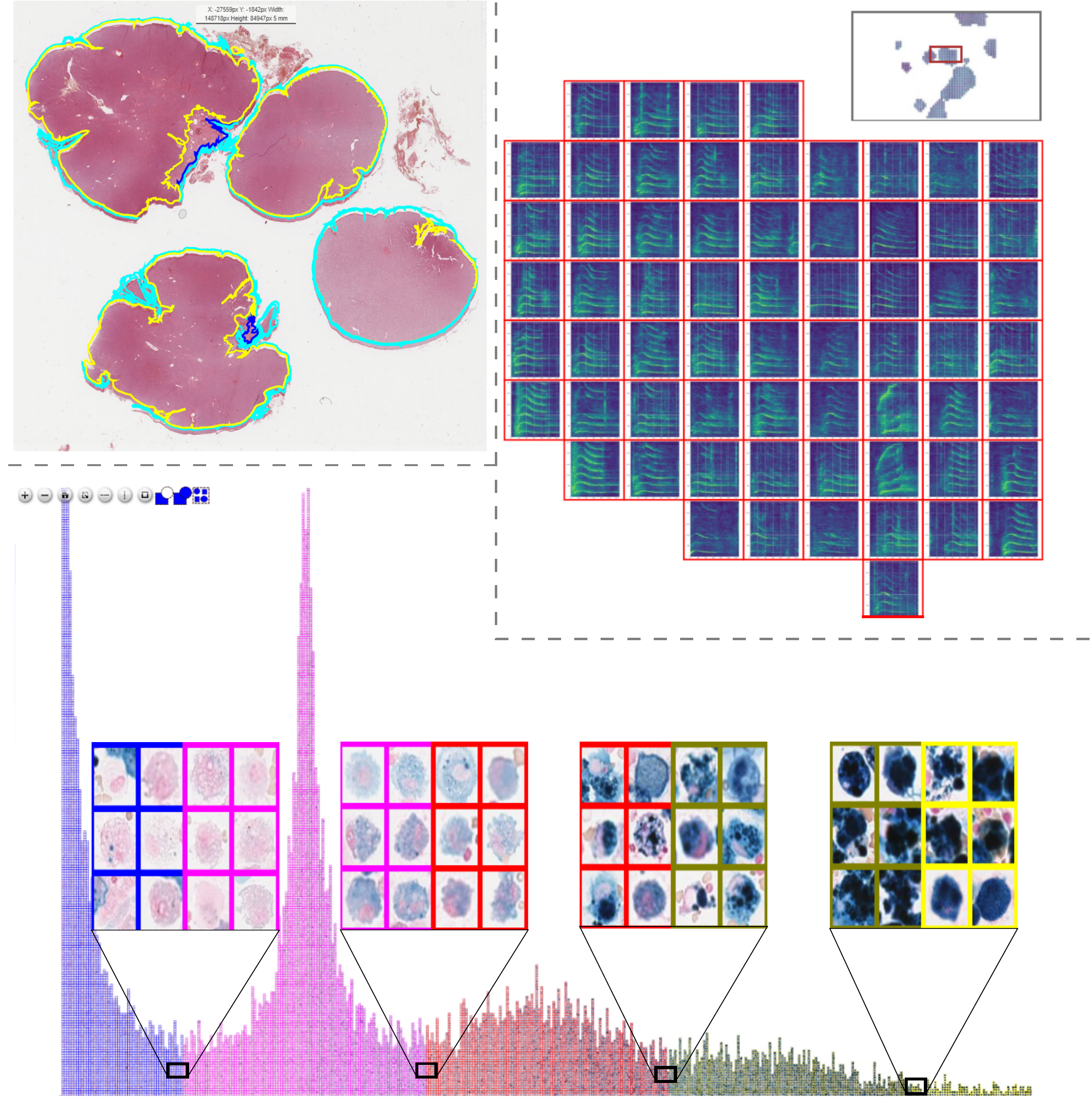}
\caption{ Top Left: Polygon annotations of a canine skin tumour tissue whole slide image. Top Right: Clustered whale sound spectroscopy images with the option to listen to the attached waveform online. Bottom: Pulmonary hemosiderophages, labelled according to their predicted class and arranged according to their predicted regression score for efficient validation by human experts.}
\label{figEXACT_Collage}
\end{figure}

\subsection*{Pathology annotation study}

In a study by Marzahl et al.~\cite{10.1007/978-3-030-59710-8_3}, EXACT was used to investigate how the efficiency of the pathology image annotation process can be increased with computer-generated pre-computed annotations. The design and results of the published study~\cite{10.1007/978-3-030-59710-8_3} showcase a prominent EXACT use case and are summarised in the following paragraphs. Ten pathologists had to perform three pathologically relevant diagnostic tasks on 20 images each, once without algorithmic support and once with algorithmic support in the form of pre-computed annotations which are visualised for the expert to review. Firstly, they had to detect mitotic figures on microscopy images. Each of the 20 images spanned ten high power fields (HPF, total area=$2.37\,mm^2$). %This is a prominent example for a rare event task, and it is one that is used in the vast majority of tumour grading schemes. 
The second task focused on performing a differential cell count in cytology of equine pulmonary fluid; a task relevant for diagnosing respiratory disease. For this, five types of visually distinguishable cells (eosinophils, mast cell, neutrophils, macrophages, lymphocytes) had to be labelled. The last task was to determine the severity of pulmonary haemorrhaging by grading the amount of breakdown products of red blood cells (hemosiderin) in alveolar macrophages according to the scoring scheme by Golde et al. \cite{golde1975occult}. 

Several EXACT features were used for this study. First of all, we used the blind annotation mode for assigning identical grading tasks to all pathology experts, which we then combined with the feature of importing pre-computed annotation for the algorithmic support. The annotation templates enabled rapid single click annotations by providing appropriate default annotation sizes for each cell type, which was particularly helpful for the equine asthma task where the different cells types have notable size differences. The systematical grading of the images was supported by the persistent screening mode plugin, which enables the expert to resume the grading process at the previously selected position on the slide at any time. %In order to estimate the impact of the EXACT server on the results, we conducted a voluntary survey at the end of the study and compared the results with a previous EIPH study \cite{marzahl2020crowd} on the same EIPH-images with the commercial software LabelBox. The grading scale ranged from 1 (very good) to 6 (insufficient). We evaluated the overall impression of the software, whether the insertion, modification or deletion of annotations was efficient and the overall effectiveness.
During the course of the study, the pathologists annotated 26,015 cells on 1200 images. The algorithmic support with EXACT lead to an increase in accuracy and a decrease of annotation time \cite{10.1007/978-3-030-59710-8_3} for all tasks. For detailed results, we kindly refer the reader to the original study. %The overall impression of the software, the persistent screening mode and the annotation effectiveness was rated with a median of two by the participating pathology experts. For the annotation mode without algorithmic support, the interaction time was significantly reduced [F(1,29)=11.23, p$<$0.01] from a mean of 105.07 to 50.43 seconds per image. \url{https://youtu.be/wjV-wHbrRjQ}
A video showcasing this study can be viewed (see Supplementary Video S1) with related source code at  \href{https://nbviewer.jupyter.org/github/ChristianMarzahl/Exact/blob/master/doc/DownloadStudyAnnotations.ipynb}{doc/DownloadStudyAnnotations.ipynb} to download the annotations. Furthermore, we added the images and ground-truth annotations from the study to the list of demo data sets which can be accessed and instantiated from the EXACT user interface.  
% Done: Show Study Video and download annotations

\subsection*{Multi-species pulmonary hemosiderophages cytology data set}

In our previous work \cite{marzahl2020deep}, 17 WSIs with 78,047 pulmonary hemosiderophages were fully annotated by a veterinary pathologist and used to develop a deep learning based object detection model. Pulmonary haemorrhage is diagnosed by performing a cytology of bronchoalveolar lavage fluid (BALF). The basis for this scoring system from Golde et al. \cite{golde1975occult} is that alveolar macrophages degrade the red blood cells into an iron-storage complex called hemosiderin. After staining the sample with Perls’ Prussian Blue or Turnbull’s Blue, the macrophages can be assigned a discrete grade from zero (low hemosiderin content) to four (high hemosiderin content). 

Building on this work, EXACT played an essential part in creating a large, fully annotated multi-species pulmonary haemorrhage data set. For this project 40 additional equine WSIs, seven feline WSIs and twelve human WSIs with evidence of chronic pulmonary haemorrhaging were annotated by expert-algorithm collaboration using EXACT and the provided object detection model. In a first step, all WSIs were annotated automatically with the deep learning model and afterwards a pathologist carefully reviewed whether all target objects were annotated. Then, the pre-computed label class was verified separately by incorporating and modifying EXACT's novel annotation map feature based on a cell-based regression approach \cite{marzahl2020deep} that reflects the continuous increase of the hemosiderin content in the target cells. This approach assigns a continuous grade between zero and four to each cell to create the annotation map for efficient manual validation (Fig. \ref{figEXACT_Collage} bottom). This annotation map orders the cells by score on the x-axis resulting in a density map of hemosiderin scores. By stacking the corresponding cell images of the same score along the y-axis, the quantity of annotated cells across the different scores is visualised (Fig. \ref{figEXACT_Collage} bottom). This enables the trained pathologist to efficiently verify the computer-generated label class by focusing on the cells which are located on the borders between two grades.
Another specialised plugin was developed to calculate the EIPH score over the current field of view in real-time (Fig. \ref{figAnnotationView}), according to Doucet et al.~\cite{doucet2002alveolar}. Code to create density maps can be accessed at \href{https://nbviewer.jupyter.org/github/ChristianMarzahl/Exact/blob/master/doc/Create\_DensityWSI-Equine.ipynb}{doc/Create\_DensityWSI-Equine.ipynb} in combination with a Supplementary Video S6. %\url{https://youtu.be/BLdX6syS_z0}.
% Done:  Screening plugin and Code für dennsity map

\subsection*{Skin tumour tissue quantification}

This ongoing project aims to segment and classify nine of the most common dog skin tumour types with deep learning algorithms. For this purpose, slides were scanned and partly annotated using SlideRunner's advanced tissue annotation tools. This project needs to synchronise the generated slides and annotations to EXACT for coordination and distribution between the participating pathology experts and computer scientists for analysis at multiple institutes and locations. SlideRunner and EXACT communicate via EXACT's REST-API to synchronise annotations, images and annotation templates (see Supplementary Video S12). EXACT's novel feature of annotation templates plays an essential role in increasing standardisation and the overall image set quality by ensuring standard annotation naming schemes and the use of polygon annotations independent of the user or user application (Fig. \ref{figEXACT_Collage} top left). While the project is actively being developed, 350 slides have already been fully annotated, resulting in 12,859 polygon annotations representing tissue layers. This indicates that a combination of online and offline tools enables fast multi-expert annotations. Code to download images, annotations and train a segmentation model can be found at \href{https://github.com/ChristianMarzahl/Exact/blob/master/doc/Segmentation.ipynb}{doc/Segmentation.ipynb} in combination with a Supplementary Video S11. %\url{https://youtu.be/AMwMvMVriGw}.
% Done:  Download and train U-Net

\subsection*{Clustering and visualisation of killer whale sounds}

While the EXACT platform is primarily developed for cooperative interdisciplinary research on microscopy images, its flexibility extends to other research areas without adaptation. We therefore showcase its use in a project that aims at deepening the understanding of killer whales \textit{(Orcinus Orca)} and their large variety of different sound types \cite{univis91996297}. In this study, EXACT is used to cluster and visualise the spectral shape of machine-pre-segmented killer whale audio samples (Fig. \ref{figEXACT_Collage} top right). Multiple EXACT features support this challenging undertaking: Firstly, the support of viewing and annotating gigapixel size images, which, in this use case, contain up to thousands of clustered spectrograms, where each spectrogram represents an individual killer whale sound. Secondly, grouped annotation assignments, which enable the user to select numerous visually grouped spectrograms simultaneously by drawing a rectangle around them in order to assign them to the same label. Finally, EXACT supports attaching media records like videos, images or sound files to the respective annotations and plays them in a web browser (Fig. \ref{figAnnotationView} left). These features enable the user to see the grouped spectrograms and additionally listen to the attached killer whale sound (see Supplementary Video S13). %\url{https://youtu.be/j0IlBcmJeLE}.
% ToDo:  Audio plugin  

%\subsection*{Confocal Laser Endomicroscopy}
% Zeitdaten 
% Sync SlideRunner
% https://www.nature.com/articles/s41598-017-12320-8

\section*{Discussion}

With the rapidly evolving digitisation of image data and the widespread use of machine learning algorithms, the need for platforms that are able to organise and display large amounts of large image data while also managing and keeping track of annotations is more crucial than ever. In this paper, we have introduced EXACT which is an open-source online platform that enables the collaborative interdisciplinary analysis of images with annotation version control. 

EXACT has proven to satisfy these requirements in several different projects ranging from collaborative tissue segmentation in the field of digital pathology to whale sound clustering. This diverse range of application represents its primary advantage. It does not only allow to extend existing offline projects with cooperation and synchronisation functions, but is also able to support researchers in various fields. Furthermore, EXACT's features provide computer scientist with version controlled annotations, advanced visualisation techniques like annotation maps or clustering, and saving artefacts from experiments like trained models. With EXACT, it is also possible to define reproducible training, validation and testing sets. 
Generally, all software solutions face the issues of support, maintenance and handling future developments. To increase the chances of turning EXACT into a successful project which offers added value for the community in the long term, EXACT will stay open-source and focus on the compatibility and synchronisation with other image analysis software.  
The flexible open-source software architecture allows for adaptation to future developments in digital pathology or other research areas. In future releases, we are planning to support a higher amount of publicly available data sets. In addition, we want to create specialised plugins exploring molecular pathology issues - an increasingly significant subdiscipline of the classic anatomical pathology. Also, valuable future extensions to EXACT include the integration of servers (like Omero \cite{allan2012omero}), which are specialised in providing microscopic images, as well as exploring options to connect EXACT with other established tools like Cytomine. Furthermore, we are investigating the integration of gamification as a promising new method to annotate data at scale. 

In summary, EXACT provides a novel feature set to boost the creation of high-quality big data sets in combination with functions to develop state-of-the-art machine learning algorithms.

\section*{Acknowledgements}

CAB gratefully acknowledges financial support received from the Dres. Jutta \& Georg Bruns-Stiftung f\"ur innovative Veterin\"armedizin.

\section*{Author contributions statement}

C. M. developed the server, created the visualisation code and wrote the main part of the manuscript. 
M. A. co-wrote the manuscript, provided code for the synchronisation with SlideRunner, provided expertise through intense discussions. 
C. A. B. co-wrote the manuscript, provided expertise through intense discussions.
J. M., J. V., C. B., C. K., K. B., R. K., A. M. provided expertise through intense discussions.

All authors contributed to the preparation of the manuscript and approved of the final manuscript for publication.

\bibliography{sample}

\begin{thebibliography}{10}
\urlstyle{rm}
\expandafter\ifx\csname url\endcsname\relax
  \def\url#1{\texttt{#1}}\fi
\expandafter\ifx\csname urlprefix\endcsname\relax\def\urlprefix{URL }\fi
\expandafter\ifx\csname doiprefix\endcsname\relax\def\doiprefix{DOI: }\fi
\providecommand{\bibinfo}[2]{#2}
\providecommand{\eprint}[2][]{\url{#2}}

\bibitem{aubreville2018sliderunner}
\bibinfo{author}{Aubreville, M.}, \bibinfo{author}{Bertram, C.},
  \bibinfo{author}{Klopfleisch, R.} \& \bibinfo{author}{Maier, A.}
\newblock \bibinfo{title}{Sliderunner - a tool for massive cell annotations in
  whole slide images}.
\newblock In \emph{\bibinfo{booktitle}{Bildverarbeitung f{\"u}r die Medizin
  2018}}, \bibinfo{pages}{309--314} (\bibinfo{publisher}{Springer},
  \bibinfo{year}{2018}).

\bibitem{hollandi2020annotatorj}
\bibinfo{author}{Hollandi, R.} \& \bibinfo{author}{Horvath, P.}
\newblock \bibinfo{journal}{\bibinfo{title}{Annotatorj: an imagej plugin to
  ease hand-annotation of cellular compartments}}.
\newblock {\emph{\JournalTitle{bioRxiv}}}  (\bibinfo{year}{2020}).

\bibitem{de2012icy}
\bibinfo{author}{De~Chaumont, F.} \emph{et~al.}
\newblock \bibinfo{journal}{\bibinfo{title}{Icy: an open bioimage informatics
  platform for extended reproducible research}}.
\newblock {\emph{\JournalTitle{Nat. Methods}}} \textbf{\bibinfo{volume}{9}},
  \bibinfo{pages}{690} (\bibinfo{year}{2012}).

\bibitem{bankhead2017qupath}
\bibinfo{author}{Bankhead, P.} \emph{et~al.}
\newblock \bibinfo{journal}{\bibinfo{title}{Qupath: Open source software for
  digital pathology image analysis}}.
\newblock {\emph{\JournalTitle{Sci. Rep.}}} \textbf{\bibinfo{volume}{7}},
  \bibinfo{pages}{1--7} (\bibinfo{year}{2017}).

\bibitem{maree2016collaborative}
\bibinfo{author}{Mar{\'e}e, R.} \emph{et~al.}
\newblock \bibinfo{journal}{\bibinfo{title}{Collaborative analysis of
  multi-gigapixel imaging data using cytomine}}.
\newblock {\emph{\JournalTitle{Bioinformatics}}} \textbf{\bibinfo{volume}{32}},
  \bibinfo{pages}{1395--1401} (\bibinfo{year}{2016}).

\bibitem{puttapirat2018openhi}
\bibinfo{author}{Puttapirat, P.} \emph{et~al.}
\newblock \bibinfo{title}{Openhi-an open source framework for annotating
  histopathological image}.
\newblock In \emph{\bibinfo{booktitle}{IEEE Int Conf Bioinformatics Biomed}},
  \bibinfo{pages}{1076--1082} (\bibinfo{organization}{IEEE},
  \bibinfo{year}{2018}).

\bibitem{obando2019icytomine}
\bibinfo{author}{Obando, D. F.~G.}, \bibinfo{author}{Mandache, D.},
  \bibinfo{author}{Olivo-Marin, J.-C.} \& \bibinfo{author}{Meas-Yedid, V.}
\newblock \bibinfo{title}{Icytomine: A user-friendly tool for integrating
  workflows on whole slide images}.
\newblock In \emph{\bibinfo{booktitle}{ECDP}}, \bibinfo{pages}{181--189}
  (\bibinfo{organization}{Springer}, \bibinfo{year}{2019}).

\bibitem{fiedler2018imagetagger}
\bibinfo{author}{Fiedler, N.}, \bibinfo{author}{Bestmann, M.} \&
  \bibinfo{author}{Hendrich, N.}
\newblock \bibinfo{title}{Imagetagger: An open source online platform for
  collaborative image labeling}.
\newblock In \emph{\bibinfo{booktitle}{Robot World Cup}},
  \bibinfo{pages}{162--169} (\bibinfo{organization}{Springer},
  \bibinfo{year}{2018}).

\bibitem{goode2013openslide}
\bibinfo{author}{Goode, A.}, \bibinfo{author}{Gilbert, B.},
  \bibinfo{author}{Harkes, J.}, \bibinfo{author}{Jukic, D.} \&
  \bibinfo{author}{Satyanarayanan, M.}
\newblock \bibinfo{journal}{\bibinfo{title}{Openslide: A vendor-neutral
  software foundation for digital pathology}}.
\newblock {\emph{\JournalTitle{J. Pathol. Inform.}}}
  \textbf{\bibinfo{volume}{4}} (\bibinfo{year}{2013}).

\bibitem{doucet2002alveolar}
\bibinfo{author}{Doucet, M.~Y.} \& \bibinfo{author}{Viel, L.}
\newblock \bibinfo{journal}{\bibinfo{title}{Alveolar macrophage graded
  hemosiderin score from bronchoalveolar lavage in horses with exercise-induced
  pulmonary hemorrhage and controls}}.
\newblock {\emph{\JournalTitle{J Vet Intern Med}}}
  \textbf{\bibinfo{volume}{16}}, \bibinfo{pages}{281--286}
  (\bibinfo{year}{2002}).

\bibitem{maier2021heidelberg}
\bibinfo{author}{Maier-Hein, L.} \emph{et~al.}
\newblock \bibinfo{journal}{\bibinfo{title}{Heidelberg colorectal data set for
  surgical data science in the sensor operating room}}.
\newblock {\emph{\JournalTitle{Scientific data}}} \textbf{\bibinfo{volume}{8}},
  \bibinfo{pages}{1--11} (\bibinfo{year}{2021}).

\bibitem{ross2021comparative}
\bibinfo{author}{Ro{\ss}, T.} \emph{et~al.}
\newblock \bibinfo{journal}{\bibinfo{title}{Comparative validation of
  multi-instance instrument segmentation in endoscopy: results of the
  robust-mis 2019 challenge}}.
\newblock {\emph{\JournalTitle{Medical image analysis}}}
  \textbf{\bibinfo{volume}{70}}, \bibinfo{pages}{101920}
  (\bibinfo{year}{2021}).

\bibitem{maaten2008visualizing}
\bibinfo{author}{Maaten, L. v.~d.} \& \bibinfo{author}{Hinton, G.}
\newblock \bibinfo{journal}{\bibinfo{title}{Visualizing data using t-sne}}.
\newblock {\emph{\JournalTitle{J MACH LEARN RES}}}
  \textbf{\bibinfo{volume}{9}}, \bibinfo{pages}{2579--2605}
  (\bibinfo{year}{2008}).

\bibitem{wold1987principal}
\bibinfo{author}{Wold, S.}, \bibinfo{author}{Esbensen, K.} \&
  \bibinfo{author}{Geladi, P.}
\newblock \bibinfo{journal}{\bibinfo{title}{Principal component analysis}}.
\newblock {\emph{\JournalTitle{CHEMOMETR INTELL LAB}}}
  \textbf{\bibinfo{volume}{2}}, \bibinfo{pages}{37--52} (\bibinfo{year}{1987}).

\bibitem{2018arXivUMAP}
\bibinfo{author}{{McInnes}, L.}, \bibinfo{author}{{Healy}, J.} \&
  \bibinfo{author}{{Melville}, J.}
\newblock \bibinfo{title}{{UMAP: Uniform Manifold Approximation and Projection
  for Dimension Reduction}} (\bibinfo{year}{2018}).
\newblock \eprint{1802.03426}.

\bibitem{orting2019survey}
\bibinfo{author}{{\O}rting, S.} \emph{et~al.}
\newblock \bibinfo{journal}{\bibinfo{title}{A survey of crowdsourcing in
  medical image analysis}}.
\newblock {\emph{\JournalTitle{arXiv preprint arXiv:1902.09159}}}
  (\bibinfo{year}{2019}).

\bibitem{marzahl2020crowd}
\bibinfo{author}{Marzahl, C.} \emph{et~al.}
\newblock \bibinfo{title}{Is crowd-algorithm collaboration an advanced
  alternative to crowd-sourcing on cytology slides?}
\newblock In \emph{\bibinfo{booktitle}{Bildverarbeitung f{\"u}r die Medizin
  2020}}, \bibinfo{pages}{26--31} (\bibinfo{publisher}{Springer},
  \bibinfo{year}{2020}).

\bibitem{10.1007/978-3-030-59710-8_3}
\bibinfo{author}{Marzahl, C.} \emph{et~al.}
\newblock \bibinfo{title}{Are fast labeling methods reliable? a case study of
  computer-aided expert annotations on microscopy slides}.
\newblock In \emph{\bibinfo{booktitle}{MICCAI}}, \bibinfo{pages}{24--32}
  (\bibinfo{publisher}{Springer International Publishing},
  \bibinfo{address}{Cham}, \bibinfo{year}{2020}).

\bibitem{golde1975occult}
\bibinfo{author}{Golde, D.~W.}, \bibinfo{author}{Drew, W.~L.},
  \bibinfo{author}{Klein, H.~Z.} \& \bibinfo{author}{{et al}}.
\newblock \bibinfo{journal}{\bibinfo{title}{Occult pulmonary haemorrhage in
  leukaemia.}}
\newblock {\emph{\JournalTitle{Br Med J}}} \textbf{\bibinfo{volume}{2}},
  \bibinfo{pages}{166--168} (\bibinfo{year}{1975}).

\bibitem{marzahl2020deep}
\bibinfo{author}{Marzahl, C.} \emph{et~al.}
\newblock \bibinfo{journal}{\bibinfo{title}{Deep learning-based quantification
  of pulmonary hemosiderophages in cytology slides}}.
\newblock {\emph{\JournalTitle{Sci. Rep.}}} \textbf{\bibinfo{volume}{10}},
  \bibinfo{pages}{1--10} (\bibinfo{year}{2020}).

\bibitem{univis91996297}
\bibinfo{author}{Bergler, C.} \emph{et~al.}
\newblock \bibinfo{journal}{\bibinfo{title}{{ORCA-SPOT: An Automatic Killer
  Whale Sound Detection Toolkit Using Deep Learning}}}.
\newblock {\emph{\JournalTitle{Sci. Rep.}}}
  \textbf{\bibinfo{volume}{10997/2019}},
  \doiprefix\url{10.1038/s41598-019-47335-w} (\bibinfo{year}{2019}).

\bibitem{allan2012omero}
\bibinfo{author}{Allan, C.} \emph{et~al.}
\newblock \bibinfo{journal}{\bibinfo{title}{Omero: flexible, model-driven data
  management for experimental biology}}.
\newblock {\emph{\JournalTitle{Nat. Methods}}} \textbf{\bibinfo{volume}{9}},
  \bibinfo{pages}{245--253} (\bibinfo{year}{2012}).

\end{thebibliography}

\section*{Supplementary information}

\textbf{Competing interests.} The authors declare no competing interests. \\
\textbf{Code availability.} 
\\ 
Server: \url{https://github.com/ChristianMarzahl/Exact} \\ 
Demo-Server: \url{https://exact.cs.fau.de/}  User: "Demo" PW: "demodemo" \\ 
REST-API Client: \url{https://github.com/ChristianMarzahl/EXACT-Sync} \\
Notebooks: \url{https://github.com/ChristianMarzahl/Exact/tree/master/doc} \\
%YouTube-Channel: \url{https://www.youtube.com/playlist?list=PLuA57h8TbJncBwIg_IJIP61EeAdYt4DSs} \\

\begin{table}[!htb]
    \centering
    \caption{EXACT documentation and references for the corresponding sections and supplementary videos.}
    \label{tabDocumentation}
    \begin{tabular}{ll}
    
    \hline 
    \multicolumn{2}{l}{\textbf{Section: Architecture}}  \\ 
    
    \href{https://github.com/ChristianMarzahl/Exact/blob/master/docker-compose.prod.yml}{docker-compose.prod.yml}           
                                               & Setup EXACT via Docker with the command  \\
                                               & \textit{docker-compose -f docker-compose.prod.yml up -d --build} \\
    %\url{https://youtu.be/-YH5cnWVrDg}         & EXACT installation guide with Docker  \\  \hline 
    %\textit{Installation.mp4}                  & EXACT installation guide with Docker  \\  \hline 
    \textit{Supplementary Video S10}           & EXACT installation guide with Docker  \\  \hline 
    
    \multicolumn{2}{l}{\textbf{Section: Application tier}}  \\ 
    \href{https://github.com/ChristianMarzahl/Exact/blob/master/exact/exact/images/models.py}{models.py}                         
                                               & Saves information to the database or file system            \\
    \href{https://github.com/ChristianMarzahl/Exact/blob/master/exact/exact/images/views.py}{views.py}                          
                                               & Creates HTML views for the presentation tier         \\
    \href{https://github.com/ChristianMarzahl/Exact/blob/master/exact/exact/images/serializers.py}{serializers.py}                    
                                               & Serialises data for the REST-API            \\
    \href{https://github.com/ChristianMarzahl/Exact/blob/master/exact/exact/images/api\_views.py}{api\_views.py}                     
                                               & Provides database CRUD operations via the REST-API            \\
    \href{https://nbviewer.jupyter.org/github/ChristianMarzahl/Exact/blob/master/doc/train\_object\_detection.ipynb}{doc/train\_object\_detection.ipynb}                                        & Code to train an object detection model via the REST-API           \\
    \href{https://github.com/ChristianMarzahl/Exact/blob/master/exact/exact/images/urls.py}{urls.py}                           
                                               & Handles the mapping between URLs and python functions \\
    \href{https://github.com/ChristianMarzahl/Exact/tree/master/exact/exact/images}{exact/images}                      
                                               & The \textit{images} module is responsible for all image-based CRUD operations  \\
    %\url{https://youtu.be/F3lV-IvT1M4}         & How to create image sets and upload images \\
    %\textit{ImageSets creation.mp4}            & How to create image sets and upload images \\
    \textit{Supplementary Video S7}            & How to create image sets and upload images \\
    %\url{https://youtu.be/VTBIyTs9lmk}         & Explain image set details \\
    %\textit{ImageSets details.mp4}             & Explain image set details \\
    \textit{Supplementary Video S8}             & Explain image set details \\
    \href{https://github.com/ChristianMarzahl/Exact/tree/master/exact/exact/annotations}{exact/annotation}                  
                                               & The \textit{annotation} module is responsible for all CRUD  \\
                                               & operations regarding annotations, verification, media files\\
                                               & and the annotation versioning system \\
    \href{https://github.com/ChristianMarzahl/Exact/tree/master/exact/plugins}{exact/plugin}                      
                                               & The \textit{plugin} module handles analysis or visualisation plugins \\
    \href{https://github.com/ChristianMarzahl/Exact/tree/master/exact/exact/users}{exact/users}                       
                                               & The \textit{users} module handles the CRUD operations for users and teams\\
    %\url{https://youtu.be/T2j5-NUhQFA}         & How to setup user access rights \\
    \textit{Supplementary Video S14}           & How to setup user access rights \\
    \href{https://github.com/ChristianMarzahl/Exact/tree/master/exact/exact/datasets}{exact/datasets}                    
                                               & The \textit{datasets} module provides features to automatically download  \\
                                               & and setup predefined data sets with their annotations\\
    \href{https://github.com/ChristianMarzahl/Exact/tree/master/exact/exact/datasets/templates}{exact/datasets/templates}          
                                               & Folder containing data set HTML templates  \\
    \href{https://github.com/ChristianMarzahl/Exact/blob/master/exact/exact/datasets/views.py}{exact/datasets/views.py}           
                                               & Implements functions to setup predefined data sets within EXACT \\
    %\url{https://youtu.be/hi23nhz0rWQ}         & How to setup and use a demo data set \\   \hline 
    %\textit{Demo Dataset.mp4}                  & How to setup and use a demo data set \\   \hline 
    \textit{Supplementary Video S5}            & How to setup and use a demo data set \\   \hline 
    
    \multicolumn{2}{l}{\textbf{Section: Inference}}  \\
    \href{https://nbviewer.jupyter.org/github/ChristianMarzahl/Exact/blob/master/doc/Inference Asthma.ipynb}{doc/Inference Asthma.ipynb}        
                                               & A REST-API inference example           \\
    %\url{https://youtu.be/xP4YAp678EM}         & Example for REST-API and JavaScript inference\\ \hline
    %\textit{Inference.mp4}                     & Example for REST-API and JavaScript inference\\ \hline
    \textit{Supplementary Video S9}            & Example for REST-API and JavaScript inference\\ \hline
                                               
    \multicolumn{2}{l}{\textbf{Section: Annotation map screening mode}}  \\ 
    \href{https://nbviewer.jupyter.org/github/ChristianMarzahl/Exact/blob/master/doc/AnnotationMap.ipynb}{doc/AnnotationMap.ipynb}           
                                               & Code to create annotation maps           \\
    %\url{https://youtu.be/GAjvOSkLW8Q}         & How to create annotation maps \\
    %\textit{Annotation Maps.mp4}               & How to create annotation maps \\
    \textit{Supplementary Video S2}               & How to create annotation maps \\
    \href{https://nbviewer.jupyter.org/github/ChristianMarzahl/Exact/blob/master/doc/ClusterCells.ipynb}{doc/ClusterCells.ipynb}            
                                               & Code to cluster Asthma cells           \\
    %\url{https://youtu.be/Wvz-Nv4dNOE}         & How to cluster Asthma cells \\  \hline 
    %\textit{Asthma Clustering.mp4}             & How to cluster Asthma cells \\  \hline 
    \textit{Supplementary Video S4}             & How to cluster Asthma cells \\  \hline 
    
    \multicolumn{2}{l}{\textbf{Section: Image set versioning and machine learning support}}  \\
    \href{https://nbviewer.jupyter.org/github/ChristianMarzahl/Exact/blob/master/doc/DownloadStudyAnnotations.ipynb}{doc/DownloadStudyAnnotations.ipynb}                                        & Code to download annotations from EXACT           \\
    %\url{https://youtu.be/WeOWxXaYc0g}         & How to create a new image set version and track changes  \\ \hline 
    %\textit{Versioning.mp4}                    & How to create a new image set version and track changes  \\ \hline 
    \textit{Supplementary Video S15}            & How to create a new image set version and track changes  \\ \hline 
    
    \multicolumn{2}{l}{\textbf{Section: Annotation templates}}  \\  
    %\url{https://youtu.be/4XdWLaqy9UA}         & How to create annotation templates with EXACT           \\ \hline 
    %\textit{Annotation Templates.mp4}          & How to create annotation templates with EXACT           \\ \hline 
    \textit{Supplementary Video S3}          & How to create annotation templates with EXACT           \\ \hline 
    
    \multicolumn{2}{l}{\textbf{Section: Pathology annotation study}}  \\  
    %\url{https://youtu.be/wjV-wHbrRjQ}         & How to download annotations\\&Explanation of parts of the annotation study           \\
    %\textit{Annotation download.mp4}           & How to download annotations and explanation of parts of the annotation study           \\
    \textit{Supplementary Video S1}           & How to download annotations and explanation of parts of the annotation study           \\
    \href{https://nbviewer.jupyter.org/github/ChristianMarzahl/Exact/blob/master/doc/DownloadStudyAnnotations.ipynb}{doc/DownloadStudyAnnotations.ipynb}& Code to download annotations via the REST-API           \\   
    
    \multicolumn{2}{l}{\textbf{Section: Multi-species pulmonary hemosiderophages cytology data set}}  \\ \hline
    %\url{https://youtu.be/BLdX6syS_z0}         & How to create density maps \\&Explanation of the EIPH plugin        \\
    %\textit{Density Maps.mp4}                  & How to create density maps and explanation of the EIPH plugin        \\
     \textit{Supplementary Video S6}            & How to create density maps and explanation of the EIPH plugin        \\
    \href{https://nbviewer.jupyter.org/github/ChristianMarzahl/Exact/blob/master/doc/Create\_DensityWSI-Equine.ipynb}{doc/Create\_DensityWSI-Equine.ipynb}& Source code for density maps           \\  \hline
    
    \multicolumn{2}{l}{\textbf{Section: Skin tumour tissue quantification}}  \\  \
    \href{https://nbviewer.jupyter.org/github/ChristianMarzahl/Exact/blob/master/doc/SyncImageAndAnnotations.ipynb}{doc/SyncImageAndAnnotations.ipynb}         
                                               & Code to synchronise between SlideRunner and EXACT          \\
    %\url{https://youtu.be/ehrfC04okyE}         & How to synchronise between SlideRunner and EXACT          \\
    %\textit{SlideRunner.mp4}                   & How to synchronise between SlideRunner and EXACT          \\
    \textit{Supplementary Video S12}           & How to synchronise between SlideRunner and EXACT          \\
    %\url{https://youtu.be/AMwMvMVriGw}         & How to segment with EXACT           \\
    %\textit{Segmentation.mp4}                  & How to segment with EXACT           \\
    \textit{Supplementary Video S11}           & How to segment with EXACT           \\
    \href{https://nbviewer.jupyter.org/github/ChristianMarzahl/Exact/blob/master/doc/Segmentation.ipynb}{doc/Segmentation.ipynb}            
                                               & Code to download information from EXACT to train a segmentation network  \\ 
    \href{https://nbviewer.jupyter.org/github/ChristianMarzahl/Exact/blob/master/doc/PatchClassifier.ipynb}{doc/PatchClassifier.ipynb}            
                                               & Code to download information from EXACT to train a patch classifier  \\ \hline
                                               
    \multicolumn{2}{l}{\textbf{Section: Clustering and visualisation of killer whale sounds}}  \\   
    %\url{https://youtu.be/j0IlBcmJeLE}         & How to perform sound clustering and visualisation           \\
    %\textit{Sound Clustering.mp4}              & How to perform sound clustering and visualisation           \\
    \textit{Supplementary Video S13}           & How to perform sound clustering and visualisation           \\
  
    \end{tabular}
\end{table}

\end{document}